\documentclass[11pt]{article}
\usepackage[top=1in, bottom=1in, left=1in, right=1in]{geometry}

\usepackage{color, amsmath, amssymb, amsbsy, amsthm, graphicx, bbm, amsfonts, bm, dsfont, courier}
\usepackage[toc,page]{appendix}
\usepackage{setspace}
\usepackage{booktabs}
\usepackage[flushleft]{threeparttable}
\usepackage{latexsym}
\usepackage[dvipsnames]{xcolor}
\usepackage{comment}
\usepackage{wrapfig}
\usepackage{graphicx}
\usepackage{graphics}
\usepackage{algorithm}
\usepackage{algorithmic}
\usepackage[colorlinks,allcolors=blue]{hyperref}
\usepackage[round]{natbib}  

\usepackage{soul}
\usepackage{threeparttable,booktabs}
\usepackage{etoolbox}
\appto\TPTnoteSettings{\footnotesize}
\usepackage{setspace}
\theoremstyle{plain}
\newtheorem{theorem}{Theorem}
\newtheorem{corollary}{Corollary}
\newtheorem{remark}{Remark}
\newtheorem{example}{Example}
\newtheorem*{assumption*}{\assumptionnumber}
\providecommand{\assumptionnumber}{}
\makeatletter
\newenvironment{assumption}[2]
{%
	\renewcommand{\assumptionnumber}{Assumption #1$\mathcal{#2}$}%
	\begin{assumption*}%
		\protected@edef\@currentlabel{#1$\mathcal{#2}$}%
	}
	{%
	\end{assumption*}
}
\makeatother

\def \bX{\mathbf{X}}
\def \bZ{\mathbf{Z}}
\def \bI{\mathbf{I}}
\def \hm{\hat{M}}
\def \hat{\widehat}
\def \tilde{\widetilde}
\def \bX{ \bm{X} }
\def \bP{ \bm{P} }
\def \bZ{ \bm{Z} }

\def \bI{ \bm{I} }

\newcommand{\cw}[1]{{\color{blue} #1}}

\usepackage{setspace}
\begin{document}

	\title{Sharp Inference on Selected Subgroups in Observational Studies}
	\author{Xinzhou Guo$^1$ \and Linqing Wei$^2$ \and Chong Wu$^3$ \and Jingshen Wang$^2$\thanks{Correspondence: jingshenwang@berkeley.edu. The research of Jingshen Wang was supported in part by the National Science Foundation (DMS 2015325).}}
	\date{%
		$^1$Harvard-MIT Center for Regulatory Science\\%
		$^2$Division of Biostatistics, UC Berkeley\\
		$^3$Department of Statistics, Florida State University\\[2ex]%
	}
	\maketitle

	\begin{abstract}
			{
			
			 In modern drug development, the broader availability of high-dimensional observational data provides opportunities for scientist to explore subgroup heterogeneity, especially when randomized clinical trials are unavailable due to cost and ethical constraints. However, a common practice that naively searches the subgroup with a high treatment level is often misleading due to the ``subgroup selection bias." More importantly, the nature of high-dimensional observational data has further exacerbated the challenge of accurately estimating the subgroup treatment effects. To resolve these issues, we provide new inferential tools based on resampling to assess the replicability of post-hoc identified subgroups from observational studies. Through careful theoretical justification and extensive simulations, we show that our proposed approach delivers asymptotically sharp confidence intervals and debiased estimates for the selected subgroup treatment effects in the presence of high-dimensional covariates. We further demonstrate the merit of the proposed methods by analyzing the UK Biobank data. The \texttt{R} package ``\texttt{debiased.subgroup}'' implementing the proposed procedures is available on GitHub.
			}
			
	\end{abstract}
	
\doublespacing	
\section{Introduction}\label{Section-introduction}

\subsection{Motivation and our contribution}

Subgroup analysis, broadly speaking, aims to uncover and confirm treatment effect heterogeneity within a population, and it has been frequently applied to randomized clinical trials (RCTs) for drug development \citep{hebert2002design, alosh2017tutorial}. In recent years, the combination of increasing availability of observational data and advancements in statistical methods and computing capacity has stimulated researchers' interest in using observational data for subgroup analysis. This has spawned the hope for deeper findings and more targeted recommendations or interventions in a wide variety of areas from education, health care, to marketing. 
Although RCTs remain the gold standard for assessing the efficacy of clinical interventions, observational data offer broader opportunities to explore subgroup heterogeneity when we cannot afford RCTs due to cost and ethical constraints.

While observational studies can be useful to identify subgroups with favorable treatment efficacy or unfavorable adverse effects, we must account for the impact of subgroup selection on any subsequent evaluation of the subgroup. We have been frequently reminded by the failure of follow-up trials to confirm a seemingly promising subgroup \citep[see][for example]{kubota2014phase}, and by the discussions in the domain science journals about the appeals and pitfalls of subgroup analysis \citep[e.g.]{petticrew2012damned}.  When subgroups are identified post hoc from RCTs, it has been recognized in the literature \citep{cook2014lessons, bornkamp2017model, Guo2019inference} that the subgroup selection bias, if not accounted for, is likely to lead to overtreatments and false discoveries. When subgroups are identified from observational studies, the problem becomes even more challenging as subgroup treatment effects need to be estimated upon adjusting for possibly high-dimensional confounders. To address these issues, in this paper, we provide new inferential tools to help assess the replicability of post-hoc identified subgroups from observational studies without having to resort to simultaneous inference methods that are often too conservative to start with.

From a statistical methodological standpoint, our proposed approach delivers asymptotically sharp confidence intervals as well as debiased estimates for the selected subgroup treatment effects in the presence of high-dimensional covariates. We break down our methodological contributions as follows: 

First, we allow well-understood debiased regression parameter estimates to be used as the building blocks for the subgroup treatment effects from observational studies, thus enabling adjustments for high-dimensional covariates in the study.  In particular, we investigate how debiased Lasso and repeated data splitting can be used in the subgroup analysis. We find that the former works well at handling a large number of subgroup-specific parameters, whereas the latter achieves better statistical efficiency at sparse models especially when the subgroup assignments are highly correlated with other covariates in the model.

Second, we propose new bootstrap-calibrated procedures for quantifying the selected subgroup treatment effect (Section \ref{Sec:debiased-boot} and \ref{Sec:rsplit-boot}) and provide a theoretical guarantee of its asymptotic validity in high dimensions (Section \ref{Section-theory}).  Most importantly, the resulting confidence intervals are asymptotically sharp in the sense that they approximate the nominal coverage probability without overshooting it (Theorem \ref{thm: theorem 1} and Corollary \ref{corollary:selected-subgroup}). While the existing statistical literature \citep{zhang2017simultaneous, dezeure2017high} has argued for the use of multiplicity adjustment to ensure post-selection inference validity in high dimensions, most recommended methods tend to sacrifice power as they aim to protect the family-wise error rates over all possible subgroups. The proposed method accounts automatically for subgroup selection bias as well as increasingly complex correlation structures among the subgroups, leading to targeted (instead of simultaneous)  statistical validity in subgroup analysis.

\subsection{Related literature}

Our paper is closely related to subgroup analysis. Here, the literature is often divided into exploratory subgroup analysis that focuses on subgroup identification \citep{shen2015inference, su2009subgroup}, and confirmatory subgroup analysis that aims to validate subgroups identified from an earlier stage \citep{jenkins2011adaptive,friede2012conditional}. These classical methods are usually made for randomized trials and are not applicable for observational studies. While some recent literature on subgroup analysis \citep{fan2017change,yang2020propensity,izem2020comparison} accounts for the pre-treatment information, their generalization to high-dimensional observation studies may require substantial modification. Beyond subgroup analysis, some methods on analyzing heterogeneous treatment effects \citep{imai2013estimating,wager2017estimation} might be also applicable for subgroup analysis in observational studies. Different from our goal of simultaneously identifying and estimating subgroup treatment effect, their goal mainly focuses on subgroup identification or making inference on a pre-defined subgroup. 

Our methodology also contributes to high-dimensional inference and post-selection inference literature, simply because accurate point estimation of treatment effects is usually not available without regularization \citep[e.g.][]{tibshirani1996regression, fan2001variable} in high dimensions.
Selective inference \citep{lee2016exact, tian2018selective} constructs exact confidence intervals for the selected regression coefficients based on Lasso conditional on the selected model, indicating that their framework tends to provide conservative confidence intervals overall. While recent developments in debiased inference \citep{zhang2014confidence, van2014asymptotically, belloni2014inference} removes the Lasso regularization bias by using an estimate of the inverse population covariance matrix, they do not address the issue of subgroup selection bias. 

Our framework is connected to the multiple comparison literature. Although several attempts have been made to address the multiple comparison issues in subgroup analysis \citep{hall2010bootstrap, fuentes2018confidence, stallard2008estimation}, those procedures are either conservative or poorly grounded. \cite{guo2020inference} propose a asymptotically sharp procedure to adjust the subgroup selection bias but it is not designed for high-dimensional observational studies. In addition, most false discovery controls that are developed for independent tests cannot be well justified for multiple analyses based on the same data set. When they protect false positive rates, they tend to be conservative too. In our proposal, we handle the dependence among tests by using the bootstrap-based calibration.

\section{Model setup and challenges} \label{Section-Model-setup-and-challenges}

In classical subgroup analysis in RCTs, a subgroup is usually defined as a subpopulation of patients decided by their baseline characteristics, and the corresponding subgroup treatment effect is the effect of a certain medical treatment for this given subpopulation. As observational studies often allow us to investigate multiple treatment strategies at the same time, subgroups can be more broadly defined. In this paper, other than simply viewing subgroups in the classical RCT setting, we also consider subgroups that are defined as subpopulations of patients that receive different treatments. We refer the variables that capture the differential subgroup treatment effects as treatment effect heterogeneity variables.  

Suppose we have a random sample of $n$ i.i.d. observations $\{(Y_i, Z_i, X_i)\}_{i=1}^{n}$, where $Y_i$ is the response variable, $Z_i$ is a $p_1$-dimensional vector of treatment effect heterogeneity variables (See Remark \ref{remark:consctruction-of-z} for its construction in different scenarios), and $X_i$ is a $p_2$-dimensional vector of covariates for the $i$-th subject.  Without loss of generality, we assume that a larger value of $\beta_i$ means a better treatment effect. We work under the high-dimensional setting where $p_1 + p_2 \gg n$.  We assume that 
    \begin{equation}\label{basic}
    Y_i=Z_i'\beta+X_i'\gamma+\varepsilon_i,\hspace{1cm} i=1,\dots,n,
    \end{equation}
    where $\beta= (\beta_1,\dots,\beta_{p_1})'$ captures the subgroup treatment effects, $\gamma$ is a sparse vector of coefficients, and $\varepsilon_i$'s are independent random errors with $\mathbb{E}(\varepsilon_i|Z_i,W_i)=0$. Define $[p_1]=\{1,\dots,p_1\}$. In practice, scientists often hope to identify the subgroup with the highest treatment level. Our goal is to find a consistent point estimate and a confidence interval with asymptotically sharp coverage probability on:
    
    \indent (i) the selected subgroup treatment effect: $\beta_{\hat{s}}, \text{   where } \hat{s} = \arg\max_{j\in[p_1]} \hat{\beta}_j$,\\
    \indent (ii) the maximal subgroup treatment effect: $\beta_{\max}=\max_{j\in[p_1]}\beta_j$,\\
    \noindent where $\hat{\beta}_i$ is an estimate of $\beta_i$ (shall be specified in Sections \ref{Sec:debiased-boot} and \ref{Sec:rsplit-boot}). 
    
    While one may debate which perspective is practically more relevant in subgroup analysis, our proposed inferential procedure works for both quantities. We will start with inference on $\beta_{\max}$ and show in Section \ref{Section-theory} (Corollary \ref{corollary:selected-subgroup}) that the same procedure works for making inference on $\beta_{\hat{s}}$. 
    
    To fully realize the challenges of making inference on $\beta_{\max}$ in high dimensions, we now provide an illustrative example. By showing this example, our take-away point for practitioners is clear: Even if $\beta$ can be estimated accurately, methods without appropriate adjustments of random selection cannot produce valid inference on $\beta_{\max}$ (or $\beta_{\hat{s}}$). 
    
\begin{wrapfigure}{r}{0.5\textwidth}
        \includegraphics[width=0.48\textwidth]{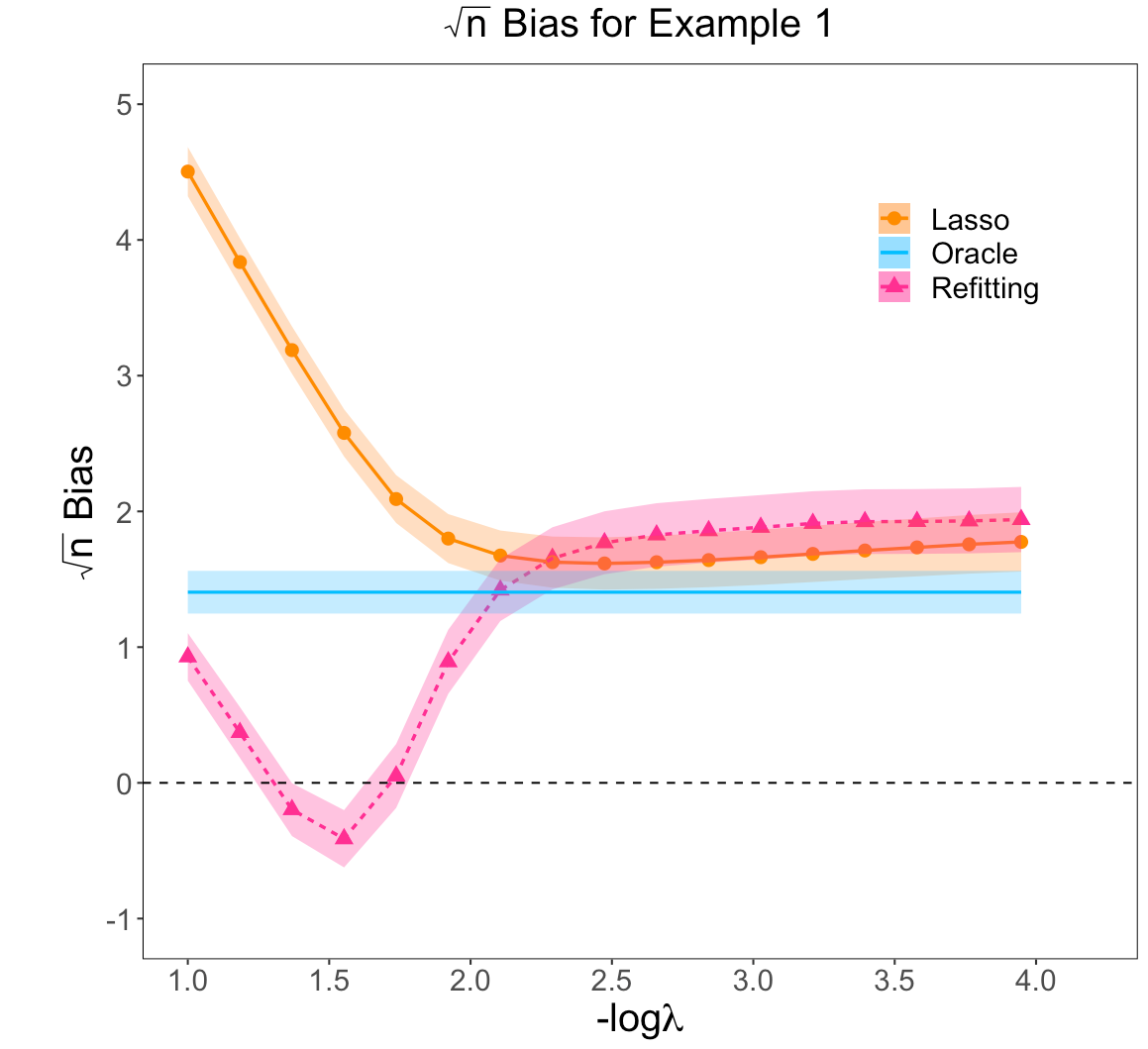}
        \caption{\small \label{fig:bias-example} Root-$n$ scaled bias along with its Monte Carlo standard error bands for Example 1.}
    \end{wrapfigure}

    \begin{example}[Selection bias and regularization bias in estimating $\beta_{\max}$]\normalfont  In practice, $\beta_{\max}$ is frequently estimated in a two-step procedure: One first obtains an estimate $\hat{\beta}$ and then estimates $\beta_{\max} $ by taking the maximum $\max\{ \hat{\beta}_1, \ldots,\hat{\beta}_{p_1}\}$. Here, we use two widely adopted procedures to estimate $\beta$ in high dimensions: (1) Lasso, which refers to simply using the estimates $\big( \hat{\beta}_{\text{Lasso}}',\hat{\gamma}_{\text{Lasso}}'\big)' $ from the $\ell_1$-penalized regression program without any adjustments, and (2) Refitted Lasso, which is obtained by refitting the linear model by the ordinary least squares (OLS) based on the covariates in the support set of $\big( \hat{\beta}_{\text{Lasso}}',\hat{\gamma}_{\text{Lasso}}'\big)' $. As a benchmark, we also report the performance of the oracle estimator $\big( \hat{\beta}_{\text{Oracle}}',\hat{\gamma}_{\text{Oracle}}'\big)' $ which pretend the true support set of $\gamma$ is known and is estimated by OLS.
    We generate Monte Carlo samples following the setup in Model \eqref{basic}. We generate $Z_{i1} = \mathbf{1}_{\{ W_{i1}>0.5 \}}$, $Z_{i2} = \mathbf{1}_{\{ W_{i2}>0.5 \}}$, $X_{ij} = W_{ij}$ for $ 3 \leq j \leq p$, and $W_i \sim N( 0, \Sigma)$ where $\Sigma = (\Sigma_{jk})_{j,k=1}^{p}$ and $\Sigma_{jk} = 0.5^{|j-k|}$ for $i=1, \ldots, n$. We set the sample size $n=100$ and the dimension $p=500$ and set the coefficients $\beta = (0.5, 0.5)'$ and $\gamma = (1, 0, \ldots)\in\mathbb{R}^{p-2}$. In Figure \ref{fig:bias-example}, we report the root-$n$ scaled bias along with its standard error bands based on 10,000 Monte Carlo samples.
    \end{example}

    From the results in Figure \ref{fig:bias-example}, we observe that all three estimators are biased towards estimating $\beta_{\max}$. Although the oracle estimator $\hat{\beta}_{\text{Oracle}}$ is an unbiased estimator of $\beta$, its maximum is usually not centered around $\beta_{\max}$.   In fact, following some explicit evidence given in \cite{nadarajah2008exact}, this simple plug-in estimate is usually biased upward. The magnitude of this bias also crucially depends on the unknown parameters $\beta$ and the covariance structure of  $\hat{\beta}_{\text{Oracle}}$. Therefore, any inference procedure  based on the naive plug-in estimates of $\beta_{\max}$ cannot be valid. This discrepancy between the naive plug-in estimate and the true maximum effect is referred to as the ``subgroup selection bias" in the literature \citep{zollner2007overcoming, cook2014lessons, bornkamp2017model, Guo2019inference}.
     The Lasso and the Refitted estimators are typically biased for $\beta$ due to regularization \citep[see][for detailed discussion]{wang2018debiased}, and they cannot avoid the selection bias issue either. The inferential framework proposed in the following sections simultaneously adjusts for the selection bias and the penalization bias, and it produces a bias-reduced estimate as well as an asymptotically sharp confidence interval of $\beta_{\max}$.

    \begin{remark}[Construction of $Z_i$]\label{remark:consctruction-of-z}
     When subgroups are defined similar to RCTs, $Z_i$ is composed of the interaction terms between the treatment indicator variable and the subgroups indicator variables. Under the Neyman-Rubin causal model \citep{neyman1923application, rubin1974estimating}, we show in Supplementary Materials (Section C) that such a construction allows us to interpret $\beta$ as the treatment effects of the considered subgroups. For now, we require that the considered subgroups do not overlap, and we show in Section \ref{Sec:Overlapping-subgroups} how our framework can be naturally extended to overlapping subgroups. 
     When subgroups are defined as subpopulations of patients with different treatments, $Z_i$ consists of $p_1$ indicator variables each of which represents a different treatment, and $\beta$ represents the corresponding treatment effects. Here, we allow each individual to receive multiple treatments.
     A similar model setup has been considered in \cite{imai2013estimating, wang2013detecting,lipkovich2017tutorial}. 
    \end{remark}

  \textit{Notation.} Denote $p = p_1 + p_2$ as the dimension of the vector $(Z_i', X_i')'$. Define the covariate matrix $\bX = (X_1, \ldots, X_n)'\in\mathbb{R}^{n\times p_2}$ and the subgroup design matrix $\bZ = (Z_1, \ldots, Z_n)'\in\mathbb{R}^{n\times p_1}$. Denote the maximal of any vector $a$ as $a_{\max} = \max\{a_j, j=1, \cdots p \} $, and denote a collection of integers from $1$ to $p$ as $[p]$.  Suppose $M$ is a subset of $\{1,\cdots,p\}$. Then for any $p$-dimensional vector $a$, $a_{M}$ is defined to be the sub-vector of $a$ indexed by $M$ and $\bX_M = \{X_{\cdot j}, j\in M\}$, where $X_{\cdot j}$ is the $j$th column of $\bX$. Define $I_{p}$ to be the $p-$dimensional identity matrix. Define $I_{M}$ and $I_Z$ to be index matrices, whose dimensions are context-specific and satisfy $(Z_i',X_i') I_{M}' = (Z_i',X_{i,M}')$ and $ I_Z(Z_i',X_{i,M}')' = Z_i' $, respectively. Define the sample covariance matrix as $\hat{\Sigma}_{M} = \frac{1}{n}\sum_{i=1}^n (Z_i',X_{i,M}')' (Z_i',X_{i,M}')$ and the projection matrix as $\mathbf{P}_{M} = \bX_M \big( \bX_M'\bX_M \big)^{-1} \bX_M'$. Lastly, $Q_Y(\tau)$ denotes the $\tau$th quantile of a random variable $Y$.

\section{Methodology}

 In this section, we propose two bootstrap-based inferential frameworks for $\beta_{\max}$.  We allow two well-understood debiased estimators to be used as building blocks: One built on the debiased Lasso \citep{zhang2014confidence, van2014asymptotically, belloni2014inference} for large $p_1$ (Section \ref{Sec:debiased-boot}), and the other one built on the repeated data splitting of \cite{wang2018debiased} for fixed $p_1$ (Section \ref{Sec:rsplit-boot}). While both procedures have appealing statistical properties, they have different strengths. On the one hand, the procedure based on the debiased Lasso allows $p_1$ to increase with $n$ and does not require the non-zero coefficients in $\gamma$ to be rather large. Therefore, it equips researchers with a flexible choice of $Z_i$. On the other hand, the procedure based on the repeated data splitting (R-Split) takes advantages of a sparse model. When subgroups are highly correlated with the covariates, it often provides a more efficient estimate of the target parameter. As this section focuses on the methodology side of each method, we delay the discussion on their comparison to Section \ref{Subsec:comparison}. 
	
	\subsection{Bootstrap assisted debiased Lasso adjustment for large $p_1$}\label{Sec:debiased-boot}
	
	When $p_1$ is a large and increases with the sample size $n$, we propose the following procedure to simultaneous address the selection bias and the penalization issues:
	
	\begin{description}
	\item[Step 1.] Construct the debiased estimator $\hat{b}$ of $\beta$ through \eqref{eq:desparse-Lasso-cts}.
	\item[Step 2.] For $j\in [p_1]$, calculate the calibration term: 
	$$\hat{c}_j(r)=(1-n^{r-0.5})(\hat{\beta}_{\max}-\hat{\beta}_{j,\text{Lasso}}),$$
	where $r$ is a positive tuning parameter between 0 and 0.5, whose order is motivated by our theoretical analysis (See Section \ref{Section-theory}). In practice, we choose $r$ via cross-validation (see Section \ref{Sec:Tuning} for implementation).
	\item[Step 3.] For $b\leftarrow 1$ to $B$ do
	\begin{enumerate}
	    \item Generate a consistent bootstrap replicate of $\hat{b}$, denoted as $\hat{b}^*$. 
	    \item Recalibrate bootstrap statistics  via
	    \begin{align*}
	        T_b^* =\underset{j\in [p_1]}{\max}(\hat{b}^*_{j}+\hat{c}_j(r))- \hat{\beta}_{\max}.
	    \end{align*}
	\end{enumerate}
	\item[Step 4.] The level-$\alpha$ one-sided confidence interval for $\beta_{\max}$ is  $[\hat{b}_{\max}-{Q}_{T^*_b}(\alpha),+\infty)$, and a bias-reduced estimate for $\beta_{\max}$ is  $\hat{b}_{\max}-\frac{1}{B}\sum_{b=1}^{B}T_b^*$.
	\end{description}
	 
	To cast some insight into the proposed framework, we comment on its methodological details from three perspectives:

	{First}, to address the penalization bias issue, we estimate $\beta$ via the de-sparsified Lasso procedure \citep{zhang2014confidence, van2014asymptotically} 
	\begin{align}\label{eq:desparse-Lasso-cts}
		\hat{b}_j = \hat{\beta}_{j,\text{Lasso}} + \tilde{Z}_{\cdot j}'\big(Y - \bZ\hat{\beta}_{\text{Lasso}} - \bX\hat{\gamma}_{\text{Lasso}}\big),\quad j=1,\ldots, p_1, 
	\end{align}
	where $\tilde{Z}_{\cdot j}$ can be viewed as a standardized residuals of $Z_{\cdot j}$ after regressing it on $\bX$, and it equals $\tilde{Z}_{\cdot j} = \frac{V_j}{V_j'Z_{\cdot j} }$ with $V_j=Z_{\cdot j}- (\bZ, \bX)_{-j}\hat{\zeta}_j$ and
	\begin{equation}\label{eq:desparse-Lasso-cts-z}
	\hat{\zeta}_j=\mathrm{\arg\min}\Big\{||Z_{\cdot j}- (\bZ, \bX)_{-j}\zeta_j||_2^2/n+\lambda_{n_j}||\zeta_j||_1: \zeta_j\in\mathbb{R}^{p-1}\Big\}.
	\end{equation}

	{Second}, in Step 3, our procedure relies on a valid bootstrap approximation of the debiased estimate $\hat{b}$. In high dimensions, we adopt the wild bootstrap approach proposed in \cite{dezeure2017high}: 
	\begin{equation}\label{dbootstrap}
	Y_i^*=Z_i'\hat{\beta}_{\text{Lasso}}+X_i'\hat{\gamma}_{\text{Lasso}}+\varepsilon^*_i,
	\hspace{1cm} i=1,\dots,n,
	\end{equation}
	where $\varepsilon_i^*=u_i\hat{\varepsilon}_i$, and $\hat{\varepsilon}_i$ is the residual from the Lasso estimate for the original sample and $u_i$ is i.i.d and independent of the data with $E[u_i]=0$, $E[u_i^2]=1$ and $E[u_i^4]<\infty$. $\hat{b}^*=(\hat{b}^*_{1},\dots,\hat{b}^*_{p_1})$ is defined as the debiased lasso estimate for $\beta$ based on the bootstrap sample $\{(Y_i^*,Z_i,X_i)\}_{i=1}^{n}$. We note that in \eqref{dbootstrap}, instead of using the debiased Lasso estimate $\hat{b}$ to generate bootstrap sample, \cite{dezeure2017high} adopts $\hat{\beta}_{\text{Lasso}}$ to ensure the bootstrap sample is indeed generated from a sparse model. In addition, this wild bootstrap procedure can incorporate heteroscedastic errors without much increase of the computational cost as the design matrix remains unchanged across different bootstrap samples.
	
	Under certain regularity conditions, \cite{dezeure2017high} have shown that the bootstrap procedure in \eqref{dbootstrap} is consistent in the sense that conditional on the data, the asymptotic distribution of $\sqrt{n}(\hat{b}^*-\hat{\beta}_{\text{Lasso}})$ is the same as the limiting distribution of the debiased Lasso estimator $\sqrt{n}(\hat{b}-\beta)$. However, similar to $\hat{b}_{\max}$ not centering around $\beta_{\max}$, $\hat{b}^*_{\max}=\underset{j\in[p_1]}{\max}\hat{b}^*_{j}$ is also not centered around $\hat{\beta}_{\max}=\underset{j\in[p_1]}{\max} \hat{\beta}_{j, \text{Lasso}}$,
	and usual high dimensional bootstrap procedures do not estimate the selection bias in $\hat{b}_{\max}$ correctly. As a result, any inference simply based on $\hat{b}^*_{\max}$ cannot be valid for making inference on $\beta_{\max}$ and appropriate adjustments to the bootstrap procedure is needed for a valid inference on $\beta_{\max}$. 
	
    {Third}, Step 3 constitutes the core of our bootstrap calibration procedure, which addresses the issue of the selection bias. There, we propose to modify $\hat{b}^*_{\max}$ as the following,
	\begin{equation}\label{modified}
	\hat{b}^*_{\mathrm{modified};\max}=\underset{j\in [p_1]}{\max}(\hat{b}^*_{j}+\hat{c}_j(r)) ,
	\end{equation}
	where $\hat{c}_j(r)=(1-n^{r-0.5})(\hat{\beta}_{\max}-\hat{\beta}_{j,\text{Lasso}})$, and $r$ 
	is a positive tuning parameter whose theoretical order is characterized in Assumptions \ref{Assumption-7} and \ref{Assumption-8}. In practice, we provide a simple cross validation procedure to adaptively choosing $r$ from data (Section \ref{Sec:Tuning}). 
	In $\hat{b}^*_{\mathrm{modified};\max}$,  clearly we make an adjustment to each debiased estimate $\hat{b}^*_{j}$ by the amount of $\hat{c}_j(r)$, which measures the distance between $\hat{\beta}_{j,\text{Lasso}}$ and the observed maximal regression coefficient $\hat{\beta}_{\max}$ from Lasso. The amount of this adjustment is large if $\hat{\beta}_{j,\text{Lasso}}$ is small, and is small if $\hat{\beta}_{j,\text{Lasso}}$ is large. 
	By adding the correction term $\hat{c}_j(r)$, we show in Section \ref{Section-theory} that the asymptotic distributions of $\sqrt{n}(\hat{b}^*_{\mathrm{modified};\max} -\hat{\beta}_{\max})$ and $\sqrt{n}(\hat{b}_{\max} - \beta_{\max})$ are  equivalent, meaning that the bootstrap modification can successfully adjust for the regularization bias and selection bias simultaneously. 
	
	\subsection{Bootstrap assisted R-Split adjustment for fixed $p_1$}\label{Sec:rsplit-boot}
	
	When $\beta$ is a low-dimensional parameter of interest, penalizing $\beta$ may not be necessary. In fact, in this case, inference on $\beta$ is frequently carried out after a sufficiently small model is selected \citep{belloni2013least, belloni2014inference}. Although any reasonable model selection procedure can be adopted, we only penalize $\gamma$ since $\beta$ is the target parameter and should always be kept in the selected model. In this case, given a properly chosen data-dependent model $\hm$, the refitted OLS estimator $\hat{\beta}_{\text{OLS}}$
	\begin{align*}
	(\hat{\beta}_{\text{OLS}}', \hat{\gamma}'_{\text{OLS}})' =\underset{\beta\in \mathbb{R}^{p_1}, \gamma\in \mathbb{R}^{|\hm|}} {\arg\min} \left\{ \frac{1}{n}\sum_{i=1}^n (Y_i - Z_i'\beta -X_{i, \hm}'\gamma)^2\right\},
	\end{align*}
	is a popular choice in practice. However, under the impact of the random model $\hm$ entering the estimation process, $\hat{\beta}_{\text{OLS}}$ is usually biased unless a perfect model is selected. As it is not the main focus of this article, we only briefly discuss the reason of this issue in this section and refer interested readers to \cite{wang2018debiased} for more examples and simulation studies. Following a similar derivation in \cite{wang2018debiased}, we decompose $\hat{\beta}_{\text{OLS}}$ as
	\begin{align*}
	\sqrt{n}(\hat{\beta}_{\text{OLS}} -\beta)=\underbrace{ I_{Z} (\hat{\Sigma}_{\hm})^{-1}\cdot \frac{1}{\sqrt{n}} \sum_{i=1}^{n} \begin{pmatrix}
		Z_i\\
		X_{i,\hm}
		\end{pmatrix} \varepsilon_i}_{=: b_{n1} \text{ (Over-fitting bias)}} + \underbrace{  \vphantom{I_{Z} (\hat{\Sigma}_{\hm})^{-1}\cdot \frac{1}{n} \sum_{i=1}^{n} \begin{pmatrix}r
		Z_i\\
		X_{i,\hm}
		\end{pmatrix} \varepsilon_i } (\bZ'(\bI- \bP_{\hm})\bZ/n)^{-1} \bZ'(\bI- \bP_{\hm}) \bX\beta/\sqrt{n}}_{=: b_{n2} \text{ (Under-fitting bias)}}.
	\end{align*}
	Due to the correlation between $\varepsilon_i$ and a data dependent model $\hm$, we typically have $\mathbb{E}(\varepsilon_i|X_{i,\hm}) \neq 0$, meaning that the first term $b_{n1}$ is roughly the mean of $n$ random variables that do not have mean zero. The second term $b_{n2}$ captures the impact of the unselected variables in $\hm$ in estimating $\beta$, and it vanishes whenever the selected model $\hm$ covers the support set of $\gamma$. To simultaneously control the over- and under-fitting biases induced by regularization procedures used in model selection, we adopt the repeated data splitting approach (R-Split) proposed by \cite{wang2018debiased} to estimate $\beta$. A detailed description of R-Split has been provided in Steps 1 and 2. Under certain regularity conditions, R-Split estimator, denoted as $\tilde{b}$, removes the over- and under-fitting bias, and it converges to a normal distribution centered around $\beta$ at a root-$n$ rate. 
	
	Note that the model selection procedure adopted in Step 1(b) can be any easily accessible procedure, but ideally, we hope this selection procedure by-pass the model selection mistakes in a high probability to avoid the risk of under-fitting. 
	Following the result given in Theorem \ref{thm: theorem 1} of \cite{wang2018debiased}, the smoothed estimator $\tilde{b}$ satisfies the following linear expansion: 
    \begin{align*}
    \tilde{b}-\beta = \Gamma_n \cdot \frac{1}{n} \sum_{i=1}^n (Z_i', X_i')' \varepsilon_i + o_p(1/\sqrt{n}),
    \end{align*}
    where $\Gamma_n$ is a $p_1$ by $p$ dimensional matrix that is independent of $\varepsilon$, and $\Gamma_n$ can be well approximated by its sample analogy $\tilde{\Gamma}_n$ defined in Step 2 in the previous section. After generating the residuals $(\varepsilon^*_{1}, \ldots, \varepsilon^*_{n})$ following \eqref{dbootstrap}, such a linear expansion allows us to quickly generate the bootstrap replicates of $\tilde{b}$ without implementing double bootstrap:
    \begin{align}\label{eq:bootstrap-rsplit}
    \tilde{b}^* =  \tilde{b}+\tilde{\Gamma}_n \cdot \frac{1}{n} \sum_{i=1}^n \begin{pmatrix}
    Z_i\\
    X_{i}
    \end{pmatrix} \varepsilon^*_i. 
    \end{align}
    We provide the theoretical justification of this bootstrap procedure in the Supplementary Material Section B. 
  
	Similar to \eqref{modified} in the previous section, with the help of a valid bootstrap procedure in replicating $\tilde{b}$, we again modify the bootstrap statistics by adding $\tilde{c}_j(r) = (1-n^{r-0.5})(\tilde{b}_{\max} - \tilde{b}_j)$ to have a consistent approximation for the distribution of $\tilde{b}_{\max}$. A detailed description of the proposed procedure is summarized below.
	
	\begin{description}
	\item[Step 1.] For $b\leftarrow 1$ to $B_1$ do
	\begin{enumerate}
	    \item Randomly split the data $\{ (Y_i, X_i, Z_i) \}_{i=1}^n$ into group  $T_1$ of size $n_1$ and group $T_2$ of size $n_2 = n-n_1$, and let   $v_{bi} = \mathbf{1}_{(i\in T_2)}$, for $i=1,\cdots,n$. 
	    \item  Select a model $ \hat{{M}}_b$ to predict $Y$  based on $T_1$.
	    \item Refit the model with the data in $T_2$ to get $(\tilde{b}_b'  ,\tilde{\gamma}_b')' = {\arg\min}\ \sum_{j\in T_2} ( Y_j - Z_j'\beta - X'_{j, 	\hat{{M}}_b }{\gamma}   )^2.$
	\end{enumerate}
	The R-Split estimate is obtained by averaging over $\tilde{b}_b$: 
	\begin{align*}
	    \widetilde{b} = \frac{1}{B_1} \sum_{b=1}^{B_1} \tilde{b}_b. 
	\end{align*}
	\item[Step 2.] For $j\in [p_1]$, calculate: 
	$$\tilde{\Gamma}_n = \frac{1}{B_1}\sum_{b=1}^{B_1}    I_z \left( \frac{1}{n_1}\sum_{i=1}^n v_{bi} \begin{pmatrix} Z_i \\
			X_{i,\hm} \end{pmatrix} (Z_i', X'_{i,\hm})' \right)^{-1}I_{\hm},\quad \tilde{c}_j(r) = (1-n^{r-0.5})(\tilde{b}_{\max} - \tilde{b}_j),$$
		where $r$ is a positive tuning parameter between 0 to 0.5. 
	\item[Step 3.] For $b\leftarrow 1$ to $B_2$ do
	\begin{enumerate}
	    \item Generate bootstrap replicate $\tilde{b}^*$ from \eqref{eq:bootstrap-rsplit}.
	    \item Recalibrate bootstrap statistics  via
	    \begin{align*}
	        T_b^* =\underset{j\in [p_1]}{\max}(\tilde{b}^*_{j}+\tilde{c}_j(r))- \tilde{b}_{\max}.
	    \end{align*}
	\end{enumerate}
	\item[Step 4.]  The level-$\alpha$ one-sided confidence interval for $\beta_{\max}$ is  $[\tilde{b}_{\max}-{Q}_{T^*_b}(\alpha),+\infty)$, and a bias-reduced estimate for $\beta_{\max}$ is $\tilde{b}_{\max}-\frac{1}{B_2}\sum_{b=1}^{B_2}{T^*_b}$. 
	\end{description}
	
\section{Theoretical investigation and a comparison}\label{Section-theory}
	
\subsection{Theoretical investigation for bootstrap assisted debiased Lasso adjustment}\label{Subsec:debiased-theory}
In this section, we discuss the theoretical properties of the bootstrap assisted debiased Lasso adjustment in detail. Define the set of indexes with the maximal regression coefficient as $H = \{ j\in [p_1]: \ \beta_j = \max_{k\in [p_1]} \beta_k \}$. To establish the asymptotic validity of the bootstrap assisted debiased Lasso procedure, under the fixed design, we work under the following assumptions:

\begin{assumption}{1}{ }\label{Assumption-1} The Lasso estimates satisfy 
	$||\hat{\beta}_{\textit{Lasso}} -\beta||_1+||\hat{\gamma}_{\textit{Lasso}}-\gamma||_1=o_p(1/\sqrt{\log(p)\log(1+p_1)}).$
\end{assumption}

\begin{assumption}{2}{ }\label{Assumption-2}
The tuning parameter for Lasso satisfies $\lambda_n\asymp\sqrt{\log(p)/n}$. 
\end{assumption}
	
\begin{assumption}{3}{ }\label{Assumption-3} The noise variables 
	$\varepsilon_1,\dots,\varepsilon_n$ are mutually independent, and satisfy (i) $\mathbb{E}(\varepsilon_i)=0$, (ii) $\mathbb{E} \sum_{i=1}^{n}\varepsilon_i^2/n=\sigma_\varepsilon^2$ and $\sigma_i^2=\mathbb{E}\varepsilon_i^2$ bounded away from zero, and (iii) $\mathbb{E}|\varepsilon_i|^4$  is bounded.
\end{assumption}

\begin{assumption}{4}{ }\label{Assumption-4} The bootstrapped Lasso estimates satisfy
$||\hat{\beta}^*_{\text{Lasso}}-\hat{\beta}_{\text{Lasso}}||_1+||\hat{\gamma}^*_{\text{Lasso}}-\hat{\gamma}_{\text{Lasso}}||_1=o_p(1/\sqrt{\log(p)\log(1+p_1)})$. 
\end{assumption}

\begin{assumption}{5}{ }\label{Assumption-5}
The treatment variables $Z_i$ and the covariates $X_i$ are bounded, for $i=1,\cdots, n$. 
\end{assumption}

\begin{assumption}{6}{ }\label{Assumption-6} For the de-sparsifed Lasso step, we require (i)
	$||V_j||_2^2/n$ is bounded away from zero, (ii) $||V_j||_4^4=o(||V_j||_2^4)$, $j\in [p_1]$, and (iii) $\max_{j\in [p_1]}||V_j||_{\infty}$ is bounded above, $\log(p_1)=o(n^{1/7})$.
\end{assumption}

\begin{assumption}{7}{ }\label{Assumption-7}
The tuning parameter and the Lasso estimate satisfy $n^r||\hat{\beta}_{\text{Lasso}}-\beta||_\infty=$ \\
$o_p(1/\sqrt{\log(1+p_1)})$.
\end{assumption}

\begin{assumption}{8}{ }\label{Assumption-8} The tuning parameter satisfies 
$\liminf_{n \to \infty}\left(n^r(\max_{j\in [p_1]}\beta_j-\max_{j\notin H}\beta_j)-\log(p_1)\right)=+\infty$.
\end{assumption}

Assumptions \ref{Assumption-1}-\ref{Assumption-6} guarantee a valid (bootstrap) debiased Lasso procedure.  \cite{dezeure2017high} requires similar conditions and identify sufficient conditions for Assumptions \ref{Assumption-1}, \ref{Assumption-2},  \ref{Assumption-4} and  \ref{Assumption-6}. We refer interested readers to \cite{dezeure2017high} for a comprehensive discussion of Assumptions \ref{Assumption-1}-\ref{Assumption-6}. 
Assumptions \ref{Assumption-7}-\ref{Assumption-8}  are required for a valid bias correction procedure for $\hat{b}_{\max}$. Intuitively, they require the tuning parameter to be neither too large (Assumption \ref{Assumption-7}) nor too small (Assumption \ref{Assumption-8}),  so that the modified bootstrap can correct the penalization bias and selection bias simultaneously. In fact, these two assumptions are immediate results of Assumptions \ref{Assumption-1}-\ref{Assumption-6} under certain conditions: if $n^r=O(\sqrt{\log(p)})$, then Assumption \ref{Assumption-1}  directly implies Assumption \ref{Assumption-7}; if $\max_{j\in [p_1]}\beta_j-\max_{j\notin H}\beta_j$ is a constant and $r\ge 1/7$, then Assumption \ref{Assumption-6} implies Assumption \ref{Assumption-8}.	 Lastly, when $p_1$ is fixed and $r\in(0,0.5)$, Assumptions \ref{Assumption-7}-\ref{Assumption-8} are automatically satisfied.

\begin{theorem}\label{thm: theorem 1}
	Under Assumptions \ref{Assumption-1}-\ref{Assumption-8}, the modified bootstrap maximal treatment effect estimator $\hat{b}^*_{\mathrm{modified};\max}$ satisfies:
	$$\sup_{c\in\mathbb{R}}|\mathbb{P}(\sqrt{n}(\hat{b}_{\max}-\beta_{\max})\le c)-\mathbb{P}^*(\sqrt{n}(\hat{b}^*_{\mathrm{modified};\max}-\hat{\beta}_{\max})\le c)|=o_p(1).$$
\end{theorem}

The proof of Theorem \ref{thm: theorem 1} is provided in Supplementary Material Section A.2. Theorem \ref{thm: theorem 1} confirms that the proposed one-sided confidence interval is asymptotically sharp in the sense that the proposed confidence interval achieves the exact nominal level as the sample size goes to infinity under mild conditions. In addition, any choice of the tuning parameter satisfying Assumptions \ref{Assumption-7}-\ref{Assumption-8} guarantees the asymptotic validity. 

Besides estimating $\beta_{\max}$, the proposed confidence interval also serves as an asymptotically sharp prediction interval for the selected treatment effect, i.e. $\beta_{\hat{s}}$. We formalize this notion in the following corollary and its proof can be found in Supplementary Material Section A.3.

\begin{corollary}[Selected subgroup with the maximal treatment effect]\label{corollary:selected-subgroup}
	Under Assumptions \ref{Assumption-1}-\ref{Assumption-8}, we have 
		$$\sup_{c\in\mathbb{R}}|\mathbb{P}(\sqrt{n}(\hat{b}_{\max}-\beta_{\hat{s}})\le c)-\mathbb{P}^*(\sqrt{n}(\hat{b}^*_{\mathrm{modified};\max}-\hat{\beta}_{\max})\le c)|=o_p(1).$$
\end{corollary}

\subsection{Theoretical investigation for bootstrap assisted R-Split adjustment}

In this section, we provide the theoretical property of the bootstrap assisted R-Split adjustment. As we work under the same assumption of the ones discussed in \cite{wang2018debiased}, we only state the theoretical conclusion to avoid redundancy.

\begin{theorem}\label{thm: theorem 2}
	Under the assumptions given in \cite{wang2018debiased}, when 
	$p_1$ is a fixed number, the modified bootstrap maximal treatment effect estimator $\tilde{b}^*_{\mathrm{modified};\max}= \underset{j\in [p_1]}{\max}(\tilde{b}^*_{j}+\tilde{c}_j(r))$ satisfies:
	$$\sup_{c\in\mathbb{R}}|\mathbb{P}(\sqrt{n}(\tilde{b}_{\max}-\beta_{\max})\le c)-\mathbb{P}^*(\sqrt{n}(\tilde{b}^*_{\mathrm{modified};\max}-\tilde{b}_{\max})\le c)|=o_p(1),$$
	and
	$$\sup_{c\in\mathbb{R}}|\mathbb{P}(\sqrt{n}(\tilde{b}_{\max}-\beta_{\hat{s}})\le c)-\mathbb{P}^*(\sqrt{n}(\tilde{b}^*_{\mathrm{modified};\max}-\tilde{b}_{\max})\le c)|=o_p(1).$$
\end{theorem}

\subsection{A power analysis: comparison between debiased Lasso and R-Split assisted bootstrap calibration}\label{Subsec:comparison}

While both procedures provided in Sections \ref{Sec:debiased-boot} and \ref{Sec:rsplit-boot} share similar statistical guarantees, R-Split has some advantages when $p_1$ is small. As discussed in \cite{wang2018debiased}, when $Z_i$ and $X_i$ are correlated, R-Split provides more efficient point estimates than the debiased Lasso method at sparse models. 
In this case, we expect that the proposed method built on R-Split provides shorter confidence intervals and hence is more powerful in detecting subgroup treatment effect heterogeneity. Moreover, R-Split does not penalize the target parameter $\beta$, meaning it is more transparent compared to the debiased Lasso method.

To verify our heuristic claim, we provide a simple simulation study which is similar to the two-stage model setup from the ones considered in \cite{wang2018debiased} and \cite{belloni2014inference}. We generate random samples from the model $Y_i=0.5+Z_i'\beta+X_i'\gamma+\varepsilon_i, \quad i=1,\dots,n$, where $\varepsilon_i$'s are i.i.d. standard normal random variables. As for the covariates in the model, we consider two cases: (1) When $Z_i$ is a vector of binary random variables, we generate $Z_i$ and $X_i$ from $Z_{ij} =  X_{i, 2j-1}+ X_{i, 2j} +\nu_{ij},  \quad j =1, \ldots, p_1$, where $\{\nu_{ij}\}_{i=1}^n$ are i.i.d. Bernoulli random variables with mean $p$, $\{ (X_{i, 2j-1},X_{i, 2j})'\}_{i=1}^n$ are i.i.d. Bernoulli random vectors that satisfy $X_{i, 2j-1} + X_{i, 2j} +\nu_{ij}\in \{0,1\}$ for all $j=1, \ldots, p_1$, and $\{ ( X_{i, p_1+1}, X_{ip})' \}_{i=1}^n$ are i.i.d. random vectors follows a multivariate normal distribution with covariance matrix $I_{p_2-p_1}$; (2) When $Z_i = (Z_{i1}, \ldots, Z_{ip_1})'$ is a vector of continuous random variables, we generate $Z_i$ and $X_i$ from 
$Z_{ij} =0.5 X_{i, 2j+3} + \frac{0.5}{\sqrt{2}} X_{i, 2j+4}  +\nu_{ij}, \quad j =1, \ldots, p_1$, 
where $\nu_i \sim N(0, 0.1\cdot I_{p_1})$ and $X_i\sim N(0,I_{p_2}) $. We fix the tuning parameter $r=0.1$ for simplicity.  We compare the power of the proposed bootstrap calibration methods based on R-Split and the debiased Lasso. For R-Split, we choose the model size via cross-validation with a minimum model size equals 3. For the debiased Lasso-based method, we present its power curves with four different choice of $\lambda$, which is the tuning parameter used to obtain $\hat{\beta}_{\text{Lasso}}$. 
Here, $\lambda.{\text{1se}}$ and $\lambda.{\text{min}}$ are the default tuning parameters provided by the \texttt{R} package \texttt{glmnet}, $\lambda_0$ is the tuning parameter adopted by the \texttt{hdi} package. To 
fully illustrate the impact of tuning parameter in the debiased
Lasso assisted bootstrap calibration, we also include a larger tuning parameter $1.1\cdot \lambda.{\text{1se}}$.

\begin{figure}[h]\label{fig:power}
\centering
\includegraphics[width=0.9\textwidth]{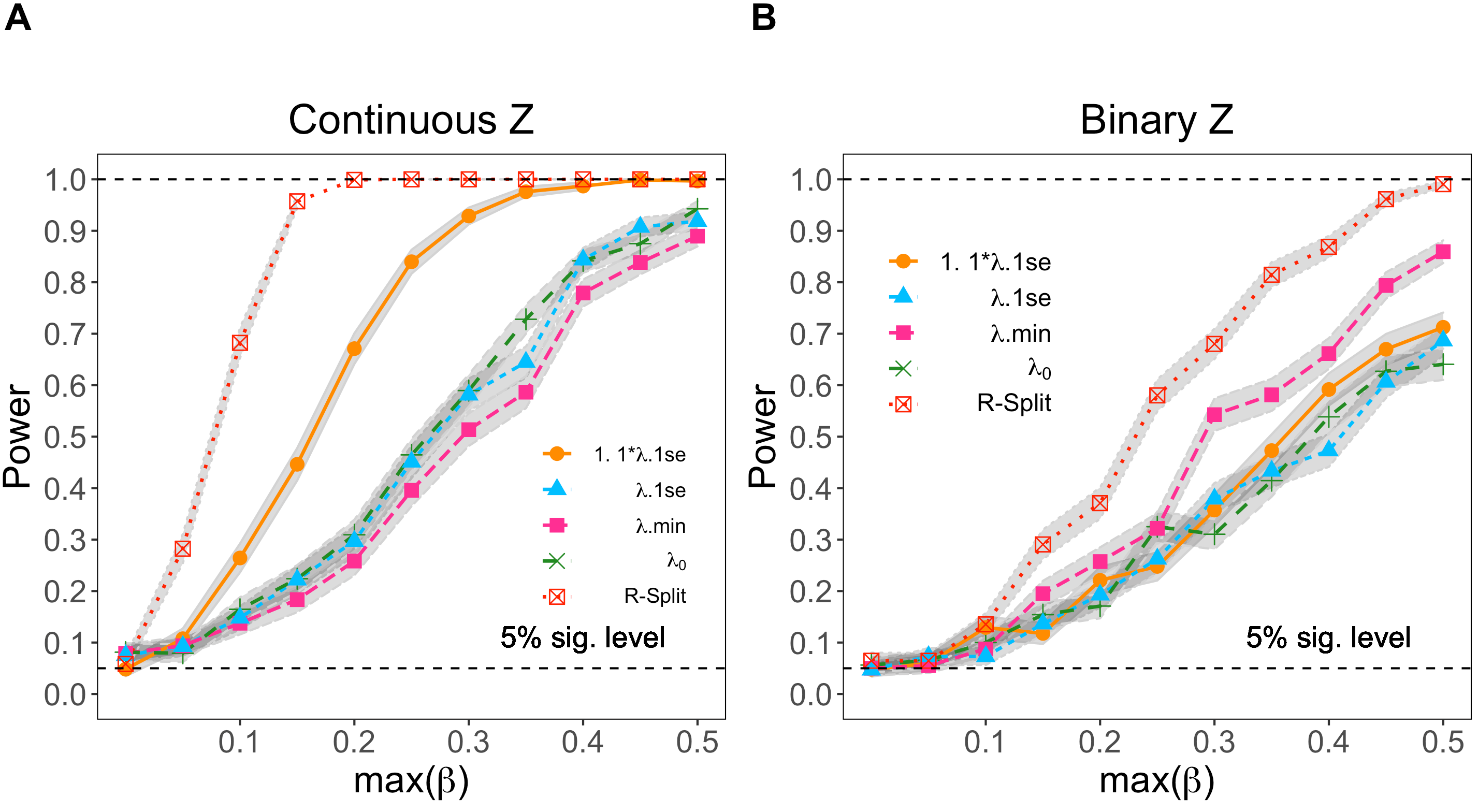}
\caption{Power comparison for R-split and the debiased Lasso assisted bootstrap calibrations. Our simulation results are evaluated through 1,000 Monte Carlo samples.}
\label{fig:power}
\end{figure}

The results in Figure \ref{fig:power} indicate that both approaches control the type-I error rate at the nominal level when $\beta_{\max} = 0$. When the covariates $X_i$ are highly correlated with $Z_i$, R-Split assisted bootstrap calibration is more powerful than the one based on the debiased Lasso. While the performance of the debiased Lasso assisted bootstrap calibration is sensitive to the choice of the tuning parameter $\lambda$, R-Split based approach is rather robust. These findings are in line with our theoretical 
analysis.

\section{Practical guide for the bootstrap calibration}

\subsection{Implementation detail and the choice of the tuning parameter $r$}\label{Sec:Tuning}
We implement the debiased Lasso-based method in Section \ref{Sec:debiased-boot} using the R package \texttt{hdi}. R-Split assisted bootstrap method in Section \ref{Sec:rsplit-boot} is implemented following the recommendation in \cite{wang2018debiased}: we select the tuning parameter (hence the model selection) in Lasso by cross-validation with a constraint on the maximum and minimum model sizes. We generate $200$ bootstrap replications for both methods and set $B_1 = 1000$ for R-Split following the recommendation in \cite{wang2018debiased}.

\begin{algorithm}
	\begin{algorithmic}[1]
		\STATE Suppose $R=\{r_1,\dots,r_m\}$ is a set of candidate tuning parameters with $r_1< \dots< r_m$ and $m$ is a finite integer. Randomly partition the data into $v$ equal-sized subsamples; 
		\vspace{0.1cm}
		\FOR{$l=1, \dots, m$}
		\vspace{0.1cm}
		\FOR{$j=1, \dots, v$}
		\vspace{0.1cm}
		\STATE \textbf{Basic setup}: use the $j$th subsample as the reference data and 
		the rest as the training data;
		
		\STATE \textbf{Bias-reduced estimator}: use the training data to obtain the bias-reduced estimator of the maximum treatment effect,  $\hat{b}_{\max,\mathrm{reduced},j}(r_l)$,  with $r_l$ as the tuning parameter;
		\FOR{$i=1, \dots, k$}
		\vspace{0.1cm}
		\STATE \textbf{Calculations on the reference data}: use the reference data to calculate the debiased lasso estimate of the $i$-th covariate of interest, $\hat{b}_{i,j}$, and its standard error $\hat{\sigma}_{i,j}$;
		\STATE \textbf{Evaluation of accuracy}: calculate $h_{i,j}(r_l)=(\hat{b}_{\max,\mathrm{reduced},j}(r_l)-\hat{b}_{i,j})^2-\hat{\sigma}_{i,j}^2$;
		\vspace{0.1cm}
		\ENDFOR
		\vspace{0.1cm}
		\ENDFOR
		\vspace{0.1cm}
		\ENDFOR
		\vspace{0.2cm}
		\STATE The tuning parameter is chosen to be  $\arg\min_{r_l} \{ \min_{i\in [k]}[\sum_{j=1}^{j=v}h_{i,j}(r_l)/v] \}$.
	\end{algorithmic}
	\caption{Cross-validated choice of the tuning parameter $r$}
	\label{algorithm:tuning}
\end{algorithm}

As for the tuning parameter $r$, we propose a data-adaptive cross-validated algorithm to select $r$ (Algorithm \ref{algorithm:tuning}). Here, the idea is to choose $r$ that minimizes the mean square error between the proposed bias reduced estimate $\hat{b}_{\max,\mathrm{reduced}}$ ($\tilde{b}_{\max,\mathrm{reduced}}$) and $\beta_{\max}$. To make this possible without knowing the true value of  $\beta_{\max}$, we provide an approximation of the mean square error that can be computed from the data (Line 8). 
The justification of this cross validation method for fixed $p_1$ can be found in \cite{guo2020inference}. 
When $p_1$ increases, we calibrate the selected tuning parameter via a simple adjustment: $r^*_{\text{cv}} = \frac{r_{\text{cv}}}{\sqrt{p_1/2}}$. Our simulation results suggest this calibration yields confidence intervals with near optimal coverage probabilities in finite samples. We leave theoretical justification for such a choice to future research. In practice, we suggest implementing Algorithm 1 via a three-fold cross-validation with a candidate set $R = \{1/3,1/6,\dots,1/30\}$.

\subsection{Overlapping subgroups}\label{Sec:Overlapping-subgroups}

   When a subgroup is defined as a subpopulation of patients decided by their baseline characteristics, following the derivation in the Supplementary Material, Model \eqref{basic} requires each subject to fall into only one of the candidate subgroups; in other words we require the considered subgroups to be non-overlapped, so that the validity of the proposed bootstrap calibration procedure is guaranteed.  Overlapping subgroups obviously induce complex correlation structure and pose difficulties in modelling, estimating, and making inference on the (selected) subgroup treatment effects. In this section, we provide a natural extension of our procedure to accommodate overlapping subgroups. Our procedure entails following steps: 
    
	\begin{description}
			\item[Step 1.] Separate the $K$ original (possibly overlapping) subgroups $S_1, \ldots, S_K$ into non-overlapping subgroups $S_1^*, \ldots, S_{p_1}^*$.  
			Let $\theta_j$ denote the subgroup treatment effects for the constructed non-overlapping subgroups for $j=1,\dots,p_1$, and let $Z_i $ denote the interaction term between the indicators of the constructed non-overlapping subgroups and the treatment indicator variable; 
			
			\item[Step 2.] Estimate the non-overlapping subgroup treatment effects either by the debiased Lasso ($\hat{\theta}$) or R-Split ($\tilde{\theta}$), and generate their corresponding bootstrap replications ($\hat{\theta}^*$ and $\tilde{\theta}^*$) via Model \eqref{basic}. For the debiased lasso procedure, we also calculate the Lasso estimator ($\hat{\theta}_{\text{Lasso}}$); 
		
			\item[Step 3.] Define a matrix $A\in\mathbb{R}^{K\times p_1}$ with 
			\begin{align*}
			A_{kj} =	\mathbb{P}\big( \text{the individual } i \text{ belongs to subgroup } S_j^*|\text{the individual } i \text{ belongs to subgroup } S_{k}  \big),
			\end{align*}
			for $k=1,\ldots,K$ and $j=1,\ldots,p_1$. In practice, $A$ is either known to us given population demographic information or could be estimated externally with some prior knowledge. 
						\item[Step 4.] Estimate the original subgroup treatment effects $\beta$ by applying a linear transformation: For the debiased Lasso procedure, we estimate the original subgroup treatment effects via {\color{black}$\hat{\beta}_{\text{Lasso}}=A\hat{\theta}_{\text{Lasso}}$, $\hat{b}=A\hat{\theta}$, and construct their bootstrap replicates via $\hat{b}^* =A\hat{\theta}^*$. Similarly, for R-Split, we obtain $\tilde{b}=A\tilde{\theta}$ and $\tilde{b}^* =A\tilde{\theta}^*$;}
		
			\item[Step 5.] Proceed with the proposed bootstrap calibration procedure discussed in Section 3 based on these reconstructed subgroup treatment effects. 
			
	\end{description}
	
    To further illustrate the proposed algorithm, we provide some explanation via a simple toy example. Suppose we consider four candidate subgroups: (1) male group $S_1$, (2) female group $S_2$, (3) young adult group $S_3$ (18 to 35 years old), and (4) senior group $S_4$ (older than 65 years old). Since these four subgroups are obviously overlapped, the coefficient of $Z_i$ in Model \eqref{basic} cannot capture the subgroup treatment effect if we naively construct $Z_i$ as the interaction term between the indicators of these four candidate subgroups and the treatment indicator variable. Step 1 of the above algorithm says we need to separate these overlapping subgroups into non-overlapping subgroups: (1) young male adult group $S_1^*$, (2) senior male group $S_2^*$, (3) young female adult group $S_3^*$, and (4) senior female group $S_4^*$. Then in Step 2, we estimate the non-overlapping subgroup treatment effects via Model \eqref{basic}. In Step 3 and 4, we reconstruct the original subgroup treatment effects of interest by applying a simple linear transformation {\color{black}$A\in\mathbb{R}^{4\times 4}$}. There, the matrix ${A}$ encodes the conditional proportion, and its component ${A}_{kj}$ represents the proportion of individuals in $S_k$ who also belong to $S_j^*$. Take the first male group for example: ${A}_{11}$ equals the proportion of young male adults in the male group, ${A}_{12} $ equals the proportion of senior males in the male group, ${A}_{13}$ and ${A}_{14} $ equal zero as there is no male in the young female adult group or the senior female group. These proportions are available to us if we have access to the population demographic information (such as those from the Census Bureau). Given these reconstructed subgroup treatment effect estimates via the linear transformation, we then proceed with the proposed bootstrap calibration procedure as usual. We provide a detailed implementation of the proposed algorithm in the Supplementary Material (Section D.1).

    Step 4 is the key element of the above algorithm, because it allows us to estimate the original subgroup treatment effects of interest by applying a simple linear transformation. As shown in the Supplementary Material (Section D.3), under Assumption \ref{Assumption-9} {\color{black}which indicates the separation is complete}, we have $\beta=A\theta$, meaning that the original true subgroup treatment effects of interest $\beta$ is connected with the non-overlapping subgroup treatment effects $\theta$ via the linear transformation. We will further argue in the Supplementary Material (Section D.2) that such complete separation required in Assumption \ref{Assumption-9} always exists. As this section focuses on the practical implementation side of our proposal, we leave the theoretical justification of the proposed algorithm to the Supplementary Material (Section D.4).

	\begin{assumption}{9}{ }\label{Assumption-9}
	The separation of $S_1, \ldots, S_K$ into non-overlapping subgroups $S_1^*, \ldots, S_{p_1}^*$ is complete in the sense that for any $j=1,\ldots,p_1$ and any $k=1,\ldots,K$,  $S^*_j\cap S_k = S^*_j \text{ or } \emptyset$. 
	\end{assumption}

\section{Simulation studies}\label{Sec:simulation-studies}
	
In this section, we consider various simulation designs to demonstrate the merit of our proposal. The main takeaway from the simulation study is as follows: When $p_1$ is a large number, the debiased Lasso assisted bootstrap calibration is more robust than the one based on R-Split.  When many of the covariates are correlated with $Z_i$ or $p_1$ is rather small, R-Split assisted bootstrap calibration is more preferable to practitioners given its high detection power. 

We generate data from the model:
\begin{align*}
Y_i=0.5+Z_i'\beta+X_i'\gamma+\varepsilon_i, \quad i=1,\dots,n, 
\end{align*}
where $n = 600$ and $\varepsilon_i$'s are i.i.d. $N(0,1)$ random variables. We consider two cases for $\beta$: (1) heterogeneous case with $\beta = (0,\ldots, 0, 1)'\in \mathbb{R}^{p_1}$, meaning that the treatment effects differ across different subgroups; and (2) spurious heterogeneous case with $\beta = (0,\ldots, 0, 0)'\in \mathbb{R}^{p_1}$, meaning that there is no significant subgroup in the population. 
We set $\gamma= (1,1,1,1,0,\dots,0)\in\mathds{R}^{p_2}$, and we leave the non-sparse case with $\gamma_j = 1/j^2, j = 1,\dots,p_2$ to Supplementary Materials (Section E). In all considered simulation designs, we set $p_1\in\{2, 6, 20\}$ and $(n, p_2) = (600, 800)$. 

As for the covariate design,  we again consider two cases similar to Section \ref{Subsec:comparison}. When $Z_i$ is a vector of binary random variables, we generate $Z_i$ and $X_i$ from 
\begin{align*}
 Z_{ij} \sim \text{Bernoulli}\Big( \frac{\exp(X_{i, 2j-1}+ X_{i, 2j})}{1+\exp(X_{i, 2j-1}+ X_{i, 2j})} \Big),  \quad j =1, \ldots, p_1, 
\end{align*} 
where $X_i\sim N(0,\Sigma) $ with $\Sigma_{ij} = 0.5^{|i-j|}$. 
When $Z_i = (Z_{i1}, \ldots, Z_{ip_1})'$ is a vector of continuous random variables, we generate $Z_i$ and $X_i$ from 
\begin{align*}
Z_{ij} =0.5 X_{i, 2j+3} + \frac{0.5}{\sqrt{2}} X_{i, 2j+4}  +\nu_{ij}, \quad j =1, \ldots, p_1,
\end{align*} 
where $\nu_i \sim N(0, I_{p_1})$ and $X_i\sim N(0,I_{p_2}) $. 

We compare the finite sample performance of the proposed R-Split and the debiased Lasso assisted bootstrap calibrations with two benchmark methods: (1) a naive method with no adjustment, which uses the estimated maximal coefficient along with its point estimates; 
and (2) the simultaneous method as discussed in \cite{dezeure2017high} and \cite{fuentes2018confidence}. For R-Split estimator, following the recommendation in \cite{wang2018debiased}, we choose the model size via cross-validation with a minimum model size equals 5. As shown in Section \ref{Subsec:comparison}, the debiased Lasso based bootstrap calibration is sensitive to the choice of $\lambda$. We report the results when it has the highest detection power: we choose $\lambda = 1.1\cdot \lambda_{\text{1se}}$ in the continuous case and $\lambda = \lambda_0$ in the binary case. We report the coverage probability, the $\sqrt{n}$ scaled Monte Carlo bias along with their standard errors based on 500 Monte Carlo samples in Table \ref{table:continuous}. 

From the results in Table \ref{table:continuous}, we observe that under spurious heterogeneity, regardless of whether $Z_i$ are binary or continuous, the method with no adjustment is clearly biased and under covered--especially when $p_1$ is rather large. Because the simultaneous method provides conservative confidence intervals, its testing power can be compromised in practice (also see real data analysis in next section). In contrast, the proposed bootstrap-assisted debiased Lasso and R-split methods perform well regardless of the underlying data generating processes. 
	
When comparing the proposed method based on R-Split with the one based on the debiased Lasso, we observe that the procedure based on the debiased Lasso is more robust than the one based on R-Split when $p_1$ is large. We conjecture that the reason is twofold. First, a large $p_1$ increases the chance of getting a singular refitting covariance matrix, as a result, R-Split can be numerically unstable if some splits do not produce feasible refitted OLS estimates. Second, the R-Split estimator may have an intractable asymptotic distribution when $p_1$ increases with $n$. In contrast, the debiased Lasso estimators are uniformly normally distributed for all components in $\beta$, suggesting that the resulting confidence intervals and tests based on the debiased Lasso are not sensitive to the change of $p_1$ as demonstrated in Section \ref{Subsec:debiased-theory}.

  \begin{table}[h!]
\begin{center}
     \caption{Simulation results \label{table:continuous} for Section \ref{Sec:simulation-studies}}
   \centering
\resizebox{\columnwidth}{!}{%
\begin{tabular}{cccccccccc}
     \hline\hline
      & & \multicolumn{3}{c}{Debiased Lasso} & & \multicolumn{3}{c}{Repeated Data Splitting} \\
     &  & Boot-Calibrated & No adjustment & Simultaneous & & Boot-Calibrated & No adjustment & Simultaneous \\\cline{3-5} \cline{7-9}
          & \multicolumn{8}{c}{ $Z_i$'s are \textbf{binary} random variables, and $\beta=(0,\dots,0,1)\in \mathbb{R}^{p_1}$ (heterogeneity)}\\\cline{2-9}
       
    $p_1=2$ &Cover     & 0.95(0.01) & 0.92(0.02)& 0.98(0.01)& & 0.96(0.01) & 0.95(0.01) & 0.97(0.01)   &  \\
    
    
    & $\sqrt{n}$Bias & -0.09(0.00)   &0.22(0.00) &-1.61(0.00)& & -0.08(0.00) & -0.09(0.00)& -1.02(0.00)&   \\[0.15cm]
    

    $p_1=6$ &Cover & 0.96(0.01) & 0.94(0.02)& 0.99(0.00)& & 0.95(0.01) & 0.94(0.01)& 0.99(0.00)&  \\
    
     
     & $\sqrt{n}$Bias &-0.07(0.00)  &0.19(0.00) &-2.35(0.00)& &  -0.10(0.00) & -0.15(0.00) &-1.91(0.00) & \\[0.15cm]
     
   
    $p_1=20$ &Cover & 0.96(0.01)  &0.96(0.01) &0.99(0.00)&  & 0.93(0.01) & 0.93(0.01)& 0.99(0.00) &  \\
    
    
    &$\sqrt{n}$Bias & -0.04(0.00)  & 0.13(0.00)&-3.16(0.00)&  & -0.12(0.00) & -0.20(0.00)& -2.69(0.00) &\\ \cline{2-9}
    
        & \multicolumn{8}{c}{$Z_i$'s are \textbf{binary} random variables, and $\beta=(0,\dots,0)\in \mathbb{R}^{p_1}$ (spurious heterogeneity)}\\\cline{2-9}
     
    $p_1=2$ &Cover & 0.96(0.01) & 0.79(0.03) &0.98(0.01)& & 0.95(0.01) & 0.87(0.02)& 0.95(0.02)& \\
    
     
     & $\sqrt{n}$Bias & -0.02(0.00)  &1.18(0.00)&-0.15(0.00)& &  0.04(0.00) & 0.91(0.00) & 0.10(0.00)  & \\[0.15cm]
     
     
    $p_1=6$ &Cover & 0.96(0.01) &0.69(0.03) &0.97(0.01)& & 0.93(0.01) & 0.77(0.03)& 0.97(0.01) &  \\
    
     
     & $\sqrt{n}$Bias & -0.03(0.00) & 1.91(0.00)&-0.25(0.00)  & & 0.14(0.00)& 1.69(0.00)  & 0.15(0.00)& \\[0.15cm]
     
 
    $p_1=20$ &Cover & 0.94(0.01)  & 0.30(0.03)&0.97(0.01)  & & 0.91(0.01) & 0.43(0.03) &0.98(0.01) & \\
    
     
     & $\sqrt{n}$Bias & -0.06(0.00) &2.72(0.00) &-0.30(0.00) & & 0.17(0.00) & 2.52(0.00) & 0.18(0.00) & \\ 
     
    \cline{2-9}
        & \multicolumn{8}{c}{$Z_i$'s are \textbf{continuous} random variables, and $\beta=(0,\dots,0,1)\in \mathbb{R}^{p_1}$ (heterogeneity)}\\\cline{2-9}
    $p_1=2$ &Cover     & 0.96(0.01) &0.90(0.01) &0.98(0.01)& & 0.96(0.01) &   0.94(0.01) &0.98(0.00) &  \\ 
    
    
    & $\sqrt{n}$Bias & -0.07(0.00)   & 0.12(0.00) &-0.43(0.00) && -0.06(0.00) & -0.09(0.00)&-0.28(0.00)  &  \\[0.15cm]

    $p_1=6$ &Cover & 0.96(0.01) & 0.91(0.01)&0.99(0.00) && 0.96(0.01) & 0.93(0.01)& 0.99(0.00)& \\
    
     
     & $\sqrt{n}$Bias & -0.05(0.00) &0.09(0.00)&-0.67(0.00)&& -0.08(0.00) &-0.11(0.00) &-0.72(0.00) &  \\[0.15cm]
   
    $p_1=20$ &Cover & 0.96(0.01) &0.94(0.01)& 0.99(0.00)&& 0.94(0.01) & 0.92(0.01)& 0.99(0.00)& \\
    
    
    &$\sqrt{n}$Bias & -0.02(0.00) &0.07(0.00)&-0.96(0.00) && -0.09(0.00) & -0.12(0.00)&-1.03(0.00) & \\
    \cline{2-9}
     & \multicolumn{8}{c}{ $Z_i$'s are \textbf{continuous} random variables, and $\beta=(0,\dots,0)\in \mathbb{R}^{p_1}$ (spurious heterogeneity)}\\\cline{2-9}
     
    $p_1=2$ &Cover & 0.94(0.01) &0.80(0.01) &0.95(0.01)&& 0.96(0.01) & 0.89(0.02)&0.97(0.02) &   \\
    
     
     & $\sqrt{n}$Bias & 0.06(0.00) &1.06(0.00) &0.02(0.00)&& 0.01(0.00) &0.24(0.00) & -0.16(0.00)&  \\[0.15cm]
     
    $p_1=6$ &Cover & 0.95(0.01) &0.84(0.02)& 0.96(0.01)&& 0.95(0.01) & 0.77(0.03)& 0.97(0.01)& \\
    
     
     & $\sqrt{n}$Bias & 0.04(0.00) & 0.77(0.00)&0.05(0.00) && 0.03(0.00)&0.72(0.00)& -0.27(0.00)  &\\[0.15cm]
 
    $p_1=20$ &Cover & 0.95(0.01) & 0.91(0.02) &0.97(0.01)&& 0.93(0.01) &0.70(0.03) &0.98(0.01) &  \\
    
     
     & $\sqrt{n}$Bias & 0.02(0.00) &0.16(0.00)&0.08(0.00)&& 0.08(0.00) &1.10(0.00) &-0.38(0.00) &\\ 
     \hline\hline
     \end{tabular}
     }
      \begin{tablenotes}\footnotesize
   \item Note: ``Cover" is the empirical coverage of the 95\% lower bound for $\beta_{\max}$ and `` $\sqrt{n}$Bias " captures the root-$n$ scaled Monte Carlo bias for estimating $\beta_{\max}$. 
     \end{tablenotes}
\end{center}
 \end{table}

		\section{Real data analysis}\label{real}
		
    In precision medicine to treat high blood pressure, lifestyle change is frequently recommended by doctors to prevent mild hypertension and to reduce the dose levels of drugs needed to control hypertension \citep{whelton2002primary}. Such lifestyle factors include, but are not limited to, sodium intake, smoking, dietary patterns, alcohol use, and physical inactivity. In addition, blood pressure is also highly heritable, and to date, more than 600 risk loci have been identified \citep{evangelou2018genetic}. 
    So far, there has been no unified answer as to (1) which genetic-risk subgroup has the most positive response to a particular lifestyle change (e.g., smoking cessation status), and (2) when subgroups are defined by different lifestyle factors, which subgroup has the best blood pressure control. In response to these questions, the identified risk loci need to be adjusted in the model-building process to avoid potential confounding issues. Our framework, which incorporates lifestyle factors as well as those high-dimensional risk loci into covariates, provides natural solutions to the questions raised above.  
    
	To demonstrate the effectiveness of our proposed method on controlling the selection and regularization bias and to cast some insights into the effects of lifestyle change on lowering systolic blood pressure, we perform cross-sectional observational studies from UK Biobank resources. The lifestyle changes we consider include smoking cessation status, three physical activity modifications, alcohol intake, and 12 diet changes. 
	
	The UK Biobank study is a long-term observational cohort study that recruited about $500,000$ individuals aged between 40 and 69 in the United Kingdom.  We choose 285,582 unrelated white British individuals, for whom systolic blood pressure, 17 lifestyle covariates, and genotype data are available. For each individual, we calculate the genetic risk scores (GRS). GRS are calculated as the weighted sum of the number of risk alleles, where the risk alleles and their weights are defined by \cite{evangelou2018genetic}. GRS profile individual-level risks of disease at birth. We focus on individuals in either high (top 0.2\% of GRS) or low (bottom 0.2\% of GRS) genetic risk groups, resulting in 857 individuals in our final analyses. We choose 0.2\% to make sure the difference between the two subgroups is significant and meaningful: the mean systolic blood pressure of the high-risk group is 144.3 mm Hg, which is 16.7 mm Hg higher than that of the low-risk group. Because blood pressure is highly heritable, 612 risk genetic variants are adjusted in our models. 
	These genetic variants are obtained from the GWAS Catalog \citep{buniello2019nhgri}, with minor allele frequency $>0.05$, and have passed standard quality control steps. Several additional covariates, including lifestyle variables, BMI, age, gender, top 40 principal components of genotype matrix, and genotype array, are adjusted as well. The response is systolic blood pressure. Our empirical comparisons are provided in Table \ref{table:realdata}. As the simultaneous method and the method with no adjustment provide similar results, we only provide their results based on the debiased Lasso adjustment.
	
    For the two-subgroup comparison, we consider low- and high-genetic risk subgroups and for whom the treatment (or lifestyle change) is smoking cessation. While the result from the naive method indicates that blood pressure control among high-genetic risk subgroup individuals is particularly reactive to smoking, the proposed methods suggest such a discovery may not be credible as the lower confidence bounds are both below zero. While both smoking and blood pressure are major risk factors for cardiovascular disease and some spurious relationships have been reported \citep{groppelli1992persistent}, current belief is that smoking has no independent effect on blood pressure control \citep{primatesta2001association} and that smoking cessation does not lower blood pressure \citep{williams20182018}. Our results not only align with the current scientific belief, but also highlight the necessity of bias correction to avoid spurious conclusions drawn from the naive method. 

    For the multiple-subgroup comparison, we compare the subgroup treatment effects of 17 lifestyle changes with a focus on the high-generic risk individuals. From Table \ref{table:realdata}, we observe that the proposed method and the naive method reach the same conclusion: the identified best treatment--moderate physical activity--shows a significant effect in controlling high blood pressure. Moderate physical activity in our analysis is defined as two or more days having 10 of  minutes or more per week of moderate physical activities like carrying light loads or cycling at normal pace. 

    In fact, this is a well-grounded conclusion \citep{williams20182018}.  Current guidelines recommend regular physical activity for preventing hypertension \citep{williams20182018} and several randomized controlled trials have further confirmed the favorable effects of physical activity on reducing blood pressure \citep{cornelissen2013exercise}. While the effect of physical activity seems small, \cw{in population level,} an systolic blood pressure decrease of 1 mm Hg decreases the risk of stroke by 5\% \citep{lawes2004blood}. 
    In both cases, because the simultaneous method is typically overly conservative and tends to lose power due to widened confidence intervals, we see that it is unable to identify any significant effect. 

     \begin{table}[!htbp]
 \begin{center}
     \caption{ Results for real data analysis  \label{table:realdata}}
     \begin{threeparttable}
     \begin{tabular}{ c c c c c}
     \hline\hline
          & Debiased & R-Split & No adjustment & Simultaneous \\ \cline{2-5}
    &\multicolumn{4}{c}{ Two subgroup comparison}\\\cline{2-5}
     Lower bound & -2.29   & -1.20 & 1.27  &  -3.78 \\
     Point estimate &  -0.13   & 1.17 & 3.26 & -0.82 \\\hline
          &\multicolumn{4}{c}{ Multiple treatment comparison }\\\cline{2-5}
      Lower bound     & 0.15 & 0.03 & 1.96 & -0.38 \\
     Point estimate & 1.75 & 1.51 & 4.44 & 1.26\\
\hline\hline
     \end{tabular}
      \begin{tablenotes}\footnotesize
   \item Note: The unit of the above results is mm Hg. 
   ``Lower bound" is the lower bound of the 95\%-coverage confidence interval.
     \end{tablenotes}
     \end{threeparttable}
     \end{center}
 \end{table}
	
\section{Conclusion}

In the presence of high-dimensional covariates from observational data, when a seemingly promising subgroup has been identified from the data, a naive estimation and inference for the identified subgroup that ignores the selection process can lead to biased and overly optimistic conclusions. 
The salient point of the present paper is that appropriate statistical analysis of the post-hoc identified subgroup treatment effect must take the selection process into account. We propose two bootstrap assisted debiasing procedures to make valid inferences on the maximal or the selected subgroup treatment effect. The two proposed methods are not only easy to implement but also simultaneously control the regularization bias and the selection bias. The resulting statistical inference is asymptotically sharp. They have different strengths: the R-Split assisted bootstrap calibration has higher detection power when the number of subgroups is small, while the debiased Lasso assisted bootstrap calibration tends to be more robust when we want to investigate multiple subgroups. Since the proposed approaches have their own strengths, it is worth exploring the merits of these two approaches and further combine them. We leave this to our future research. 

\subsubsection*{Software and reproducibility}
\texttt{R} code for the proposed procedures can be found in the package ``\texttt{debiased.subgroup}'' that is publicly available at \texttt{https://github.com/WaverlyWei/debiased.subgroup}. Simulation examples can be reproduced by running examples in the \texttt{R} package.

\subsubsection*{Acknowledgement}
	We thank the individuals involved in the UK Biobank for their participation and the research teams for their work on collecting, processing, and sharing these datasets. This research has been conducted using the UK Biobank Resource (application number 48240), subject to a data transfer agreement. We are also grateful for the helpful discussions with Peng Ding, Avi Feller, Xuming He, Xinwei Ma, Gongjun Xu, and the paticipants of the UC Berkeley Causal Inference Reading group. 
	
	\bibliographystyle{apa}
	\bibliography{SharpInference}
    \end{document}